\documentclass[aps,prl,reprint,superscriptaddress,showpacs,floatfix,twocolumn]{revtex4-1}
\usepackage{graphicx}
\usepackage{dcolumn}
\usepackage{bm}

\begin{document}


\title{Interplay of electron-electron and electron-phonon interactions in the low temperature phase of 1$\emph{T}$-TaS$_{2}$}


\author{Doohee Cho}
\affiliation{Center for Artificial Low Dimensional Electronic Systems, Institute for Basic Science (IBS), 77 Cheongam-Ro, Pohang 790-784, Republic of Korea}
\affiliation{Department of Physics, Pohang University of Science and Technology (POSTECH), Pohang 790-784, Republic of Korea}
\author{Yong-Heum Cho}
\affiliation{Department of Physics, Pohang University of Science and Technology (POSTECH), Pohang 790-784, Republic of Korea}
\affiliation{Laboratory for Pohang Emergent Materials, Pohang University of Science and Technology, Pohang 790-784, Korea}
\author{Sang-Wook Cheong}
\affiliation{Laboratory for Pohang Emergent Materials, Pohang University of Science and Technology, Pohang 790-784, Korea}
\affiliation{Rutgers Center for Emergent Materials and Department of Physics and Astronomy, Rutgers University, Piscataway, New Jersey 08854, USA}
\author{Ki-Seok Kim}
\affiliation{Department of Physics, Pohang University of Science and Technology (POSTECH), Pohang 790-784, Republic of Korea}
\author{Han Woong Yeom}
\email{yeom@postech.ac.kr}
\affiliation{Center for Artificial Low Dimensional Electronic Systems, Institute for Basic Science (IBS), 77 Cheongam-Ro, Pohang 790-784, Republic of Korea}
\affiliation{Department of Physics, Pohang University of Science and Technology (POSTECH), Pohang 790-784, Republic of Korea}


\date{\today}

\begin{abstract}
We investigate the interplay of the electron-electron and electron-phonon interactions in the electronic structure of an exotic insulating state in the layered dichalcogenide 1$\emph{T}$-TaS$_{2}$, where the charge-density-wave (CDW) order coexists with a Mott correlation gap. Scanning tunneling microscopy and spectroscopy measurements with high spatial and energy resolution determine unambiguously the CDW and the Mott gap as 0.20 -- 0.24 eV and 0.32 eV, respectively, through the real space electron phases measured across the multiply formed energy gaps. An unusual local reduction of the Mott gap is observed on the defect site, which indicates the renormalization of the on-site Coulomb interaction by the electron-phonon coupling as predicted by the Hubbard-Holstein model. The Mott-gap renormalization provides new insight into the disorder-induced quasi-metallic phases of 1$\emph{T}$-TaS$_{2}$.
\end{abstract}



\maketitle


Metal-insulator transitions in low dimensional condensed matter systems are driven by various interactions between relevant degrees of freedom, such as spin, charge, orbital, and lattice. For example, charge density waves (CDW)~\cite{gruner1988dynamics} and Mott insulators~\cite{mott1968metal} are paradigmatic examples of electron-phonon ($\emph{e-ph}$) and electron-electron ($\emph{e-e}$) couplings, respectively. These couplings are often entangled to yield exotic states of electrons as recently discussed for high-temperature superconductors~\cite{hayward2014angular}. On the other hand, the interplay of $\emph{e-ph}$ and $\emph{e-e}$ couplings has been discussed for a while for the CDW transition~\cite{wilson1975charge} in a layered transition metal dichalcogenide of 1$\emph{T}$-TaS$_{2}$~\cite{fazekas1979electrical}. Upon decreasing temperature, it undergoes a series of transitions from a metallic phase through incommensurate and nearly commensurate CDW to commensurate (C)-CDW ~\cite{wilson1975charge}. The ordered CDW superstructures could be directly observed using scanning tunneling microscopy (STM) in real space~\cite{wu1990direct,kim1994observation}. The C-CDW superstructure splits the broad metallic band into a manifold of narrow subbands but cannot account for the most important gap opening at the Fermi level (E$_{F}$) with one half-filled band left~\cite{smith1985band}. The insulating gap was then explained by introducing the on-site Coulomb repulsion, which is enhanced by the  substantial narrowing of the band width due to the CDW formation~\cite{fazekas1979electrical}. Thus, the multiple gap structure in this system itself is a hallmark of the interplay between $\emph{e-ph}$ and $\emph{e-e}$ couplings.

However, while this material has been investigated for a long time, the multiple gap structure is not fully characterized yet by experiments. The gap structure was investigated by angle resolved photoemission spectroscopy (ARPES)~\cite{clerc2006lattice}, inverse-ARPES~\cite{sato2014conduction} and scanning tunneling spectroscopy (STS)~\cite{kim1994observation} below the critical temperature (T$_{c}$$\sim$180 K). ARPES probed only the occupied part of the gap and inverse-ARPES for the empty part did not provide proper energy resolution. The previous STS study only focused on the gap at E$_F$ without any clear information on the multiple gaps. Moreover, these spectroscopic works reported substantially larger gaps than those measured by optical and transport measurements~\cite{barker1975infrared,kobayashi1979anomalously,gasparov2002phonon}. Considering the fact that the multiple gap structure is crucially important to understand not only the ground state but also the statically or dynamically excited states with the anomalous metallic property and the superconductivity of this system~\cite{sipos2008mott,ang2012real,ang2013superconductivity,lahoud2014emergence,stojchevska2014ultrafast}, the experimental establishment of the energy gap structure in its ground state is imperative. It is also important to trace how the Mott or CDW gap changes upon external perturbations or internal fluctuations in this respect. Moreover, generally speaking, beyond the gap structure, little is known about the physical consequences of the coupled $\emph{e-e}$ and $\emph{e-ph}$ interactions in this interesting system.

In this Letter, we report the STS results with substantially improved resolution, which identifies and quantifies clearly the multiple gaps. The origins of different gaps are unambiguously determined by the real space phases of relevant electrons. Furthermore, we unveil for the first time the local reduction of the Mott gap at the defect site, which is presumably renormalized by the local variation of the $\emph{e-ph}$ coupling. This manifests the interplay of $\emph{e-ph}$ and $\emph{e-e}$ couplings within the Hubbard-Holstein model and has important implications for the emergence of intriguing metallic C-CDW phases with intrinsic or extrinsic disorders~\cite{isa2002mid,lahoud2014emergence}.

We performed STM and STS measurements with an ultrahigh vacuum cryogenic STM (Unisoku, Japan) at 78 K. The real space electron density as a function of energy was acquired via spatially resolved (SR) differential conductance (dI/dV) measurements; the normalized tunneling conductance [(dI/dV)/(I/V)] is proportional to the local density of states (LDOS)~\cite{chen1993introduction}. The lock-in technique with a bias modulation of V$_{rms}$=20 mV, $\emph{f}$=1 kHz was applied to acquire dI/dV. 1$\emph{T}$-TaS$_{2}$ single crystals were grown by the chemical vapor transport method with iodine as a transport agent. The crystals were cleaved in high vacuum before cooling down.
\begin{figure}
\includegraphics{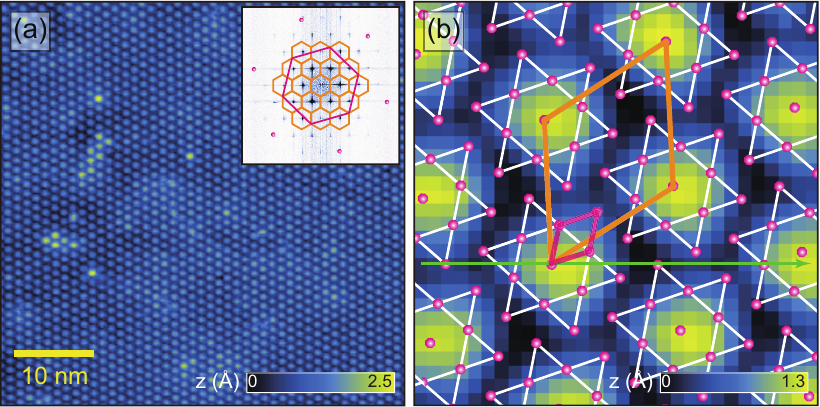}
\caption{\label{fig:epsart}(color online) (a) STM image (tunneling current I$_{t}$=100 pA, sample bias V$_{s}$=+0.2 V and 50$\times$50 nm$^{2}$) of 1$\emph{T}$-TaS$_{2}$ at 78 K displaying long range ordered CDWs. The inset shows its corresponding 2D Fourier transform. Orange and purple hexagons (parallelograms in Fig.~\ref{fig1}(b)) indicate the 1st Brillouin zone (unit cells in real space) of $\emph{p}$($\sqrt{13}$$\times$$\sqrt{13}$)R13.9$^{\circ}$ C-CDW and 1$\times$1 normal states, respectively. (b) STM image (3.5$\times$3.5 nm$^2$ and 30$\times$30 pixel) acquired during dI/dV measurements (I$_{t}$=100 pA and V$_{S}$=--1.20 V). The superimposed sketches highlight the units of the CDW.}\label{fig1}
\end{figure}

Below the transition temperature of $\sim$180 K, the CDW forms a long-range ordered superstructure~\cite{kim1994observation,clerc2006lattice} as shown in the STM topographic image of Fig.~\ref{fig1}(a). Its Fourier transformation [the inset of Fig.~\ref{fig1}(a)] shows strong peaks and their higher orders with a wavelength of $\sqrt{13}$a$_{0}$$\sim$12.1 {\AA}, which are rotated clockwise by 13.9$^{\circ}$ with respect to the $1\times1$ lattice of 1$\emph{T}$-TaS$_{2}$. The zoom-in image of Fig.~\ref{fig1}(b) shows how the CDW maxima match with the underlaying Ta lattice. The lattice is distorted into the unit of 13 Ta atoms with 12 atoms in the shape of a star of David displaced toward the center atom.
\begin{figure}
\includegraphics{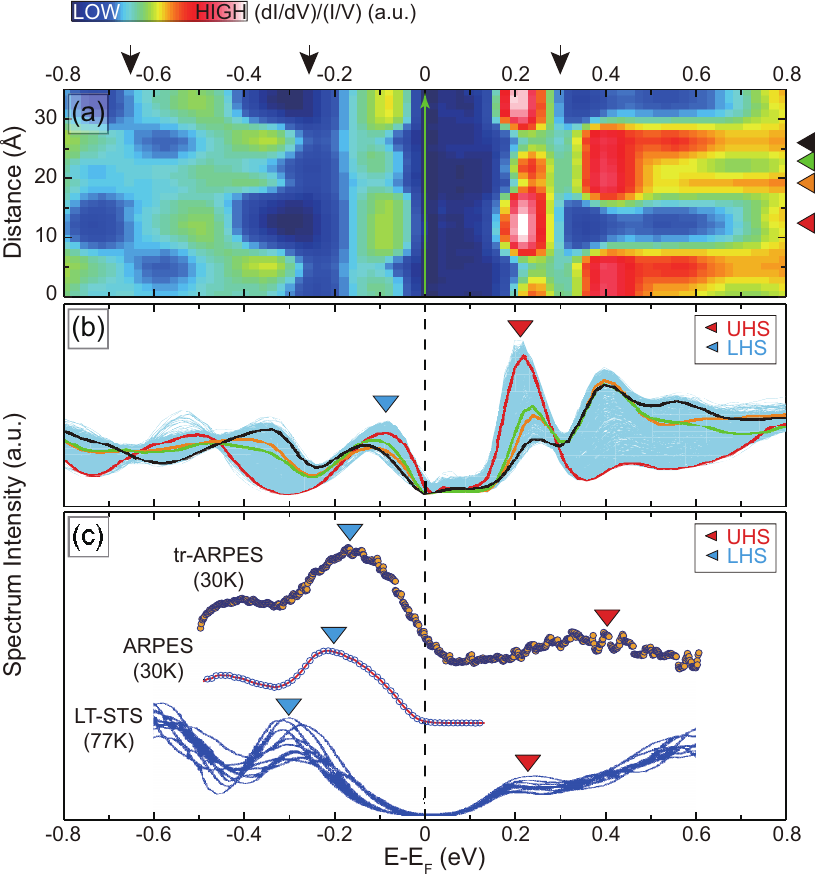}
\caption{(color online)(a) Normalized dI/dV spectrum acquired along the green arrow in Fig. 1(b). (b) A collection of 900 normalized dI/dV spectra over the area shown in Fig.~\ref{fig1}(b). The highlighted spectra (red, orange, green, and black) are obtained at the marked positions in Fig.~\ref{fig2}(a). (c) Previously reported spectra, acquired by STS~\cite{kim1994observation}, ARPES and tr-ARPES~\cite{perfetti2008femtosecond} for the C-CDW phase. The red (blue) triangle indicates the U(L)HS.}\label{fig2}
\end{figure}

In order to detail the electronic structure of the C-CDW phase, SR-STS measurements were performed simultaneously with the topographic imaging of Fig.~\ref{fig1}(b). A particular line scan along the green arrow in Fig.~\ref{fig1}(b) is shown in Fig.~\ref{fig2}(a) and a few typical normalized conductance curves in Fig.~\ref{fig2}(b).
The opening of the energy gap close to E$_{F}$ is clearly resolved. A gap size is unambiguously determined by the LDOS peaks at +0.22 and --0.10 eV as 0.32 eV. They would correspond to the coherent peaks of upper and lower Hubbard states [U(L)HS]. More clear evidence of the Mott gap will be discussed below. Depending on samples and tip conditions, the positions of coherent peaks vary within a few tens of meV~\cite{kim1994observation}. This can be related with the spatial inhomogeneity of the doping level induced by intrinsic impurities. However, the size of the correlation gap is consistent with 320 meV and E$_{F}$ is always closer to the LHS. In addition to the gap at E$_{F}$, three other gap-like features are identified around +0.30, --0.25 and --0.65 eV (marked by arrows in Fig.~\ref{fig2}), which divide the measured spectra into five subbands.

The previous spectroscopic results summarized in Fig.~\ref{fig2}(c) consistently indicated the insulating property of the C-CDW phase of 1$\emph{T}$-TaS$_{2}$~\cite{kim1994observation,perfetti2008femtosecond}. However, the estimated gap size ranges from 500 to 600 meV, which is substantially larger than the present result. This variation can stem largely from experimental limitations in the energy range or the resolution. Considering the broad line shape and the limited resolution, the ARPES results~\cite{zwick1998spectral,clerc2006lattice,perfetti2008femtosecond} seem roughly consistent with the present results for the LHS. The even more limited resolution and sensitivity are likely to prevent properly identifying the UHS in the inverse-ARPES~\cite{sato2014conduction} and time-resolved ARPES works~\cite{perfetti2008femtosecond}. The line shape of the very early STS measurement suggests that it could not resolve the first and the second peak of filled states in our STS spectra~\cite{kim1994observation}. The suppression of the spectral intensity near E$_{F}$, especially for the LHS peak, is noticeable in the previous measurement, which is highly possible for a low conductance tip condition. Furthermore, this pioneering work did not address the multiple gap structure discussed in the following. On the other hand the optical conductivity works measured a much smaller band gap of 100 -- 200 meV~\cite{barker1975infrared,gasparov2002phonon}. If one considers the finite spectral width of the U(L)HS peak in STS, the present measurement agrees very well with this measurement, establishing firmly the standard for the Mott gap of this material. The quantitative calculations taking into account of the correlation effect consistently reproduced the gap size of about 200 meV with the on-site Coulomb energy of 2 -- 4 eV ~\cite{yu2014orbital,darancet2014three}.

\begin{figure}
\includegraphics{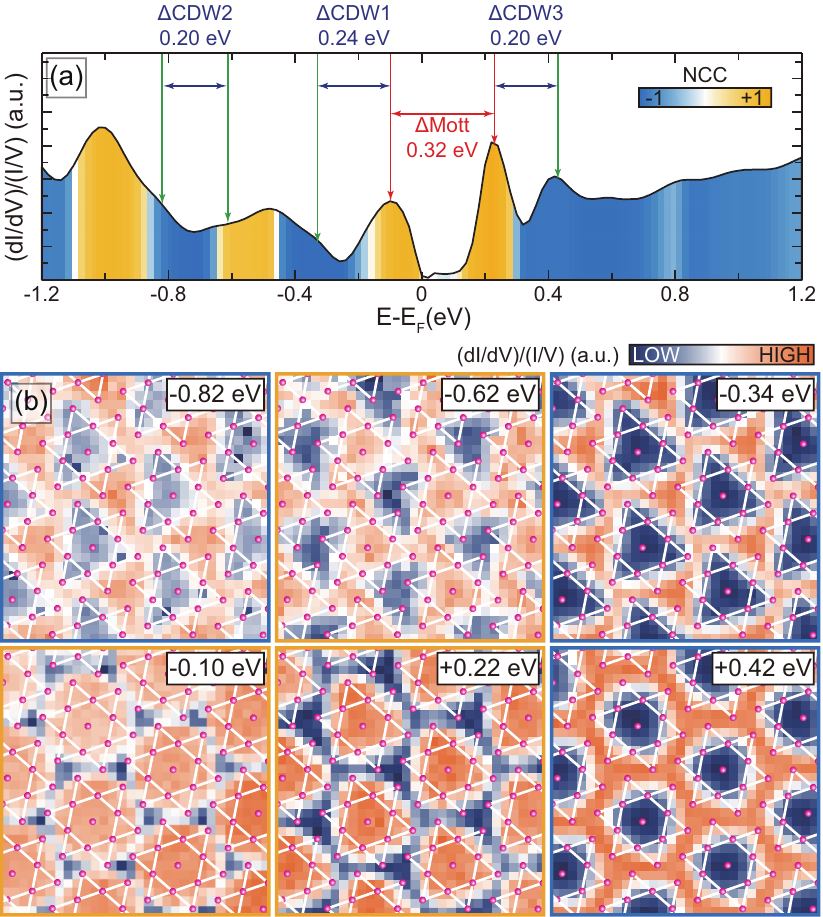}
\caption{(color online)(a) Spatially averaged STS spectrum. The color scale indicates the normalized correlation coefficient (NCC) as referenced by the image at +0.22 eV. The green arrows denote the onset of the subband manifolds. (b) LDOS maps of the states marked with arrows in Fig. 3(a) on a 3.5$\times$3.5 nm$^2$ area. The frame colors of the maps represent their NCC.}\label{fig3}
\end{figure}

As mentioned above, in addition to the Mott gap at E$_F$, we found a few other gaps and their distinct characteristics can be shown clearly through the real space phase of relevant electrons. The phase of the 2D real space electron density was analyzed as referenced by that of the UHS [the (dI/dV)/(I/V) map  at +0.22 eV shown in Fig.~\ref{fig3}(b)]. The apparent phase difference can be noticed immediately in the maps taken at different biases in the figure; while the map for UHS has bright maxima at the centers of stars of David, the maps at --0.82, --0.34, and +0.42 eV have dark minima. In order to quantify the phase difference, for example, between two states $\alpha_{ij}$ and $\beta_{ij}$, we use the normalized correlation coefficient (NCC) between two corresponding images, NCC=$\left\langle(\alpha_{ij}-\mu_{\alpha})(\beta_{ij}-\mu_{\beta}) \right\rangle$/($\sigma_{\alpha}\sigma_{\beta}$), where $\mu$ and $\sigma$ are averages and standard deviations of the images~\cite{dai2014microscopic}. The spatially averaged STS spectrum in Fig.~\ref{fig3}(a) was colored with the value of NCC obtained, showing how the relative real space phase changes for different electronic states. In particular, we notice that the phase completely flips at energies around --1.10, --0.85, --0.65, --0.45, --0.18, and +0.30 eV, which are obviously associated with the multiple spectral gaps as given in the figure. This is in clear contrast with the gap at E$_F$, crossing which the electron are apparently in-phase in real space. That is, the present result makes clear the distinction of the gap at and away from E$_F$ experimentally.

The phase flip of the real space electron density is naturally expected for any band gap formed by a periodic lattice distortion and thus can straightforwardly be related to the CDW in the present system~\cite{arguello2014visualizing}. From the onsets of the subbands identified by the detailed inspection of the spatially resolved STS spectra shown in Fig.~\ref{fig2}, we can quantify the gap sizes as 0.22$\pm$0.02 eV; three CDW gaps, two in the occupied and the other in the unoccupied states, have a similar size. This manifests further the common origin of these gaps, the periodic potential provided by CDW and the lattice distortion. Note that the CDW gap positions in the energy agree well with the previous band structure calculations ~\cite{smith1985band,rossnagel2006spin} and the ARPES measurements ~\cite{zwick1998spectral,clerc2006lattice,sohrt2014fd}. The reported gap sizes of 100--150 meV and 100--200 meV (only the occupied state gaps), respectively, are in reasonable agreement with the present quantification.



Beyond the distinct character of the Mott and CDW gaps, the real space electron densities of UHS and LHS show the clear particle-hole asymmetry; the former is more localized at the center Ta atom than the latter [Fig.~\ref{fig3}(b)]. This wave function characteristics seems consistent with the STS spectra where the UHS has a narrower band width (larger spectral peak intensity) than the LHS. This asymmetry in 1$\emph{T}$-TaS$_{2}$ has not been reported so far in this system and its origin is not clear at this point.

\begin{figure}
\includegraphics{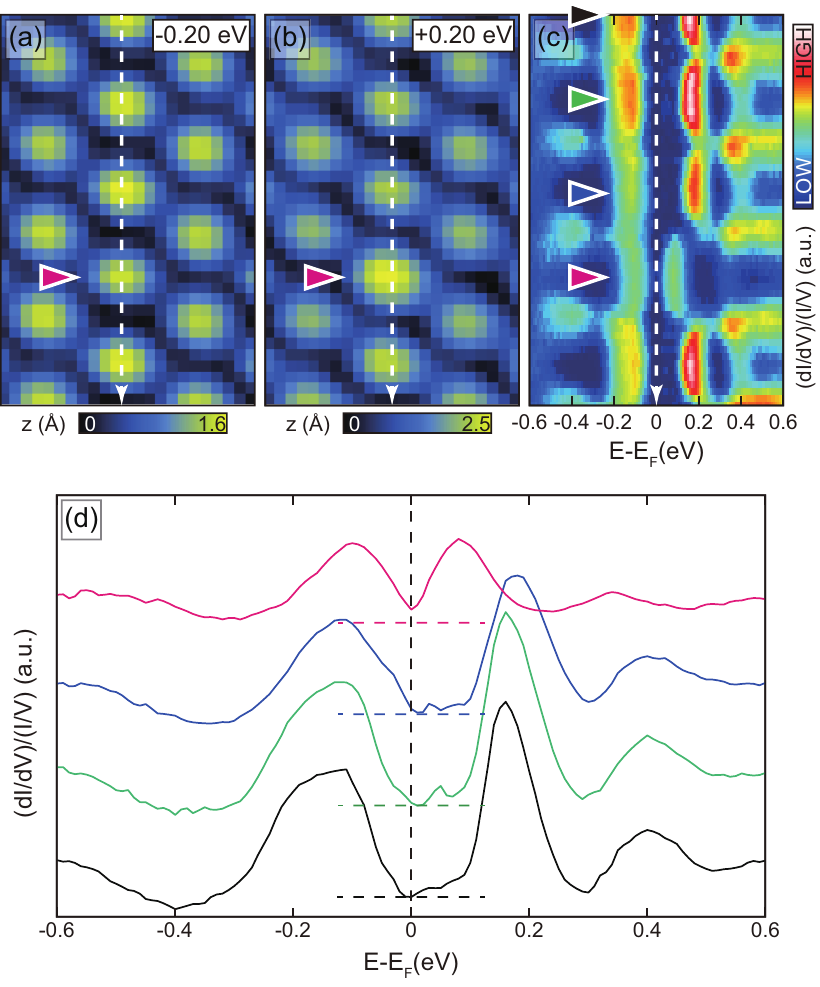}
\caption{(color online) Filled (a) and empty (b) state STM images (I$_{t}$=100 pA, V$_{s}$=$\mp$0.20 eV and 3.5$\times$5.5 nm$^2$) of an impurity on the 1$\emph{T}$-TaS$_{2}$ surface, marked by the triangles. (c) Normalized differential conductance map measured along the white dashed arrow in Fig.~\ref{fig4}(a) and (b). (d) Representative normalized dI/dV spectra acquired at the CDW crests denoted by triangles in Fig.~\ref{fig4}(c).}\label{fig4}
\end{figure}

For disordered 1$\emph{T}$-TaS$_{2}$, intriguing metallic behaviors were reported previously~\cite{zwick1998spectral,xu2010superconducting}. This metallic state was explained by the screening of the $\emph{e-e}$ Coulomb interaction through charge carriers of the discommensurations induced by the defects. This sounds similar to the nearly-commensurate CDW phase found at higher temperatures. Note that the effect of disorders on the conductivity of a correlated system can also be influenced by the aforementioned particle-hole asymmetry in theory~\cite{denteneer2001particle}. More importantly, no direct and clear spectroscopic information is available how a defect affects the Mott and CDW gaps in the present system. We thus focus on the most popular point defect in this system; the brighter CDW maxima in the STM image of Fig.~\ref{fig1}(a). Figures ~\ref{fig4}(a) and ~\ref{fig4}(b) (I$_{t}$=100 pA and V$_{s}$=$\mp$0.20 eV) show a reduced and enhanced contrast with respect to the normal CDW maximum on this defect at filled and empty state images, respectively. Within a defect concentration of our sample, they do not destroy the long or short range CDW ordering at all. The SR-STS measurement across this defect shows unexpectedly that the Mott gap is reduced considerably from 320 to 180 meV only on this site while the CDW gaps are preserved. The asymmetric U(L)HS peak becomes symmetric. Considering the change of the E$_F$ position into the center of the gap compared to the surrounding, this defect is thought to be slightly negatively charged. The local reduction of the Mott gap has never been reported previously and is sharply contrasting with the known behavior of charged impurities in a conventional Mott insulator; the spectral weight of the U(L)HS is transferred partially to a adjacent gap state~\cite{eskes1991anomalous,okada2013imaging,ye2013visualizing}. Since the CDW order and the CDW gap are preserved, the local change of the Mott gap cannot be explained in relation to CDW.

However, the reduction of the gap size can be explained by the other aspect of the interplay between $\emph{e-e}$ and $\emph{e-ph}$ couplings within the Hubbard-Hostein model. In this model, the effective $\emph{e-e}$ coupling is renormalized by the interaction between electrons and optical phonons as given by $U_{eff}=U_{e-e}-2g^{2}/\omega_{0}$ ($\omega_{0}$ denotes the optical phonon frequency and $g$ the $\emph{e-ph}$ coupling constant)~\cite{berger1995two}. The previous optical conductance measurements evidenced the optical phonons at the relevant energy range of 25 meV, which also exhibit the temperature-dependent softening~\cite{gasparov2002phonon}. This strongly suggests that the optical phonons can also be tuned by the local structural deformations. Our scenario is that the defect-induced local structural deformation reduces the optical phonon frequency and in turn leads to the reduced effective interaction $U_{eff}$. The reduced $U_{eff}$ would make the Mott gap smaller and the finite width of the U(L)HS drives the system marginally metallic as shown in Fig.~\ref{fig4}. This result bears important implications for the appearance of the quasi-metallic behavior in the C-CDW phase of disordered 1$\emph{T}$-TaS$_{2}$~\cite{isa2002mid,lahoud2014emergence}. 


In summary, we have demonstrated the interplay between $\emph{e-ph}$ and $\emph{e-e}$ coupling in the electronic structure of the low temperature ground state of 1$\emph{T}$-TaS$_{2}$ using spatially resolved scanning tunneling spectroscopy. The multiple gap structure consisting of Mott and CDW gaps, the primary hallmark of the interplay, is well quantified and their distinct origins are unambiguously visualized through the real space phases of the relevant electrons. The impurity-induced local reduction of the Mott gap is revealed and interpreted to reflect the effect of the local $\emph{e-ph}$ interaction within the Hubbard-Holstein model, which is another hallmark of the interplay.

\begin{acknowledgments}
This work was supported by Institute for Basic Science (Grant No. IBS-R015-D1). YHC and SWC are partially supported by the Max Planck POSTECH/KOREA
Research Initiative Program (Grant No. 2011-0031558) through NRF of Korea funded by MEST.
\end{acknowledgments}


\end{document}